\documentclass{article}
\usepackage{graphics,amsmath,epsfig,latexsym,amssymb,cite,graphicx,a4}
\newlength{\indentedwidth} \newdimen\mathindent
\indentedwidth=\mathindent

\def\L{\Lambda}
\def\o{\omega}
\def\d{\delta}

\def\s{\sigma}
\def\t{\tau}
\def\ve{\varepsilon}

\def\End{\mbox{End }}

\newcommand{\ov}[1]{\ensuremath{\overline{#1}}}

\DeclareMathAlphabet{\mathpzc}{OT1}{pzc}{m}{it}
\begin{document}
\vskip 0.5cm
\begin{center} {\Large \bf Universal spectral parameter-dependent Lax operators for the Drinfeld double of the dihedral group $D_3$}
\end{center}
\centerline{K.A. Dancer%
\footnote{\tt dancer@maths.uq.edu.au} 
and J. Links%
\footnote{\tt jrl@maths.uq.edu.au}}
~~\\
\centerline{\sl 
 Centre for Mathematical Physics, School of Physical Sciences, }
\centerline{\sl The University of Queensland, Brisbane 4072,
 Australia.} 
\vskip 0.9cm
\begin{abstract}
\vskip0.15cm
\noindent
Two universal spectral parameter-dependent Lax operators are presented in terms of the elements of the Drinfeld double $D(D_3)$ 
of the dihedral group $D_3$. Applying representations of $D(D_3)$ to these yields matrix solutions of the Yang-Baxter equation with spectral parameter.   
\end{abstract}

\section{Introduction}

The Yang--Baxter equation arises in several areas of mathematical physics. In the framework of quasi-triangular Hopf algebras, Drinfeld's double construction \cite{d86} provides a systematic means of constructing solutions for the Yang--Baxter equation. 
The double construction produces a canonical element, known as the universal $R$-matrix, which solves the Yang--Baxter equation algebraically. For applications to areas such as  statistical mechanics lattice models \cite{b82,yg89} and integrable quantum systems \cite{ji90,s92}, solutions which depend on a spectral parameter are necessary (which we hereafter refer to as {\it parametric} solutions in contrast to {\it constant} solutions). For some years quantum algebras (deformations of the universal enveloping algebras of Lie algebras) were a fertile field of study as an application of the Drinfeld construction, whereby many parametric solutions of the Yang-Baxter equation were obtained. Specifically, through use of an evaluation homomorphism from an affine quantum algebra to a non-affine subalgebra it is in principle possible to construct algebraic parametric solutions of the Yang-Baxter equation. Such a solution is a powerful tool in constructing a general class of matrix solutions by applying different algebra representations to the algebraic solution. However this procedure is technically challenging and only a few cases for low rank algebras have been made fully explicit \cite{zg94,bgz95}. The more common practice was to start with a constant matrix solution of the Yang--Baxter equation and then determine an analogous parameteric form, a process which is commonly known as {\it Baxterisation} \cite{j90,bk91,cgx91,zgb91,dgz94,g94,dgz96,bm01}. An alternative approach was to study the parametric Lax operator       
which is defined in the tensor product of an algebra and one of its representations. However the only instances of quantum algebras for which a solution for the parametric  Lax operator is known are $U_q(gl(n))$ \cite{j86} and its ${\mathbb Z}_2$-graded generalisation 
$U_q(gl(m|n))$ \cite{z92}.  

While quantum algebras, being a seminal class of quasi-triangular Hopf algebras, generated substantial activity as a tool for solving the parametric Yang--Baxter equation there has not been as much interest in the study of other quasi-triangular Hopf algebras for this purpose. One such example is the class of Drinfeld doubles of finite group algebras. These algebras have been studied as appropriate symmetry algebras for the description of non-abelian anyons where the conjugacy classes and centraliser subgroups of the finite group label generalised notions of the magnetic and electric charges respectively \cite{wildb98}. The study of non-abelian anyons is currently active in relation to  proposals for topological quantum computation \cite{k03,m03,nssfd}. A recent analysis of an integrable one-dimensional quantum system for non-abelian Fibonacci anyons with three-body interactions can be found in \cite{ftltkwf07}. We have previously shown that by using the Drinfeld double of the simplest non-abelian finite group $D_3$, yielding the quasi-triangular Hopf algebra denoted $D(D_3)$, there is a solution of the Yang-Baxter equation which naturally leads to an integrable model of non-abelian anyons with two-body interactions \cite{dil06}. Moreover this solution can be extended to the Drinfeld doubles of general dihedral groups $D(D_n)$ \cite{fdil}. Here we continue work in this direction by presenting two explicit universal parametric Lax operators for  $D(D_3)$, one associated with a two-dimensional representation and the other for a three-dimensional representation. This result suggests the existence of a universal parametric $R$-matrix associated with $D(D_3)$, and motivates future study. We expect that a comprehensive understanding of the solutions of the Yang--Baxter equation with symmetries given by the Drinfeld doubles of finite groups will ultimately provide important insights into interacting systems with non-abelian anyonic degrees of freedom.    
 
\section{The Drinfeld double of the dihedral group $D_3$}

The dihedral group $D_3$ (also known as the symmetric group $S_3$) represents the symmetries of a triangle, 
and has two generators $\sigma, \tau$ satisfying:

$$ \sigma^3 = e,\; \tau^2 = e,\; \tau \sigma = \sigma^{2} \tau $$

\noindent where $e$ denotes the identity.  The Drinfeld double \cite{d86} of $D_3$, denoted $D(D_3)$, has basis $$\{ gh^*| g,h \in D_3\},$$ where $g$ are the group elements and $g^*$ are dual elements. This gives an algebra of dimension 36. Multiplication of dual elements is given by 

$$g^* h^* = \delta(g,h) g^*$$

\noindent where $\d$ is the usual delta function.  The products $h^*g$ are computed using

$$h^*g = g(g^{-1} h g)^* \qquad g h^* = (ghg^{-1})^* g.$$

The algebra $D(D_3)$ becomes a Hopf algebra by imposing the following coproduct, antipode and counit respectively:

\begin{align*}
&\Delta(gh^*) = \sum_{k \in G} g (k^{-1}h)^* \otimes g k^* = \sum_{k \in G} gk^* \otimes (h k^{-1})^*, \\
&S(gh^*) = (h^{-1})^*g^{-1} = g^{-1} (g h^{-1}g^{-1})^*, \\
&\ve(gh^*)= \d(h,e), \forall g,h \in D_3.
\end{align*}
Note that we identify $g \ve$ with $g$ and $e g^*$ with $g^*$ for all $g \in G$.
The universal $R$-matrix is given by 

$$ \mathcal{R} = \sum_{g \in D_3} g \otimes g^* .$$

\noindent This can easily be shown to satisfy the defining relations for a quasi-triangular Hopf algebra:
\begin{eqnarray}
\mathcal{R}\ov{\Delta}(a)&=&\ov{\Delta}^T(a)\mathcal{R}, \quad \forall\,a\in D(G), \label{qt1}\\
(\ov{\Delta}\otimes {\rm id}) \mathcal{R}&=&\mathcal{R}_{13}\mathcal{R}_{23},   \label{qt2}\\
({\rm id}\otimes \ov{\Delta})\mathcal{R}&=& \mathcal{R}_{13}\mathcal{R}_{12}, \label{qt3}
\end{eqnarray}  
where $\ov{\Delta}^T$ is the opposite coproduct
$$ \ov{\Delta}^T(gh^*) = \sum_{k \in G} gk^* \otimes g (k^{-1}h)^* = \sum_{k \in G}
g(kh^{-1})^* \otimes   gk^*. $$ 
It follows from the relations (\ref{qt1},\ref{qt2},\ref{qt3}) that $\mathcal{R}$ is a solution of the constant 
Yang-Baxter equation, which is given by

$$\mathcal{R}_{12} \mathcal{R}_{13} \mathcal{R}_{23} = \mathcal{R}_{23} \mathcal{R}_{13} \mathcal{R}_{12}.$$

\noindent The Yang-Baxter equation operates on the three-fold tensor product space $D(D_3)^{\otimes 3}$, and $\mathcal{R}_{ij}$ indicates that $\mathcal{R}$ is acting on the $i$th and $j$th spaces.  For example, $\mathcal{R}_{12} = \mathcal{R} \otimes I$, where $I$ denotes the identity operator.

Throughout this paper we also use the following variant of the Yang-Baxter equation involving a spectral parameter:  

$$R_{12}(x/y) R_{13}(x) R_{23}(y) =  R_{23}(y) R_{13}(x) R_{12}(x/y).$$

\noindent Here the Yang-Baxter equation acts on $\End (V \otimes V \otimes V)$ where $V$ is a vector space 
and $R(x) \in \End(V \otimes V)$.  We refer to a solution to this equation as an $R$-matrix, and in particular we use the solutions given in \cite{dil06}.

Given an $R$-matrix $R(x) \in \End (V\otimes V)$, we define $\mathfrak{L}(x) \in \End V \otimes D(D_3)$ to be a 
{\it universal Lax operator} if $\mathfrak{L}(x)$ satisfies the following equation:

\begin{equation} \label{Lax}
R_{12}(x/y) \mathfrak{L}_{13}(x) \mathfrak{L}_{23}(y) =  \mathfrak{L}_{23}(y) \mathfrak{L}_{13}(x) R_{12}(x/y).
\end{equation}

\noindent By applying a representation to the third space we obtain the well-known $RLL=LLR$ equation with spectral parameter, which acts on $\End{V \otimes V \otimes W}$ and is given by 

\begin{equation}
R_{12}(x/y) L_{13}(x) L_{23}(y) =  L_{23}(y) L_{13}(x) R_{12}(x/y). \label{RLL}
\end{equation}

\noindent Here $R(x) \in \End (V \otimes V)$ is an $R$-matrix and $L \in \End (V \otimes W)$ is known as a (matrix) Lax operator.


\subsection{Representation theory of $D(D_3)$}

The Hopf algebra $D(D_3)$ has two 1-dimensional irreducible representations (irreps), four 2-dimensional irreps and two 3-dimensional irreps.  They are given by:

\begin{itemize}
\item \underline{1-dimensional irreps} \\

$\pi_{(1,\pm)} = 1, \quad \pi_{(1,\pm)}(\t) = \pm 1, \quad \pi_{(1,\pm)}(g^*) = \delta(g,e).$

\item \underline{2-dimensional irreps} \\
Set $\o$ to be the cube root of unity $\o = e^{2\pi i/3}$. Then 

$$\pi_{(2,e)}(\s) = \begin{pmatrix}
                    \omega & 0  \\ 0 & \omega^{-1}
                 \end{pmatrix}, \quad
\pi_{(2,e)}(\tau) = \begin{pmatrix}
                 0 & 1 \\ 1 & 0 
               \end{pmatrix}, \quad 
\pi_{(2,e)}(g^*) = \delta(g, e) I_2$$

\noindent and 

$$\pi_{(2,i)}(\s) = \begin{pmatrix}
                    \omega^i & 0  \\ 0 & \omega^{-i}
                 \end{pmatrix}, \quad
\pi_{(2,i)}(\tau) = \begin{pmatrix}
                 0 & 1 \\ 1 & 0 
               \end{pmatrix}, \quad 
\pi_{(2,i)}(g^*) = \left[ \delta(g, \s)+ \delta(g,\s^{-1}) \right] I_2$$

\noindent for $i = 0,1,2.$  Throughout $I_n$ will denote the $n$-dimensional identity matrix.

\item \underline{3-dimensional irreps} \\

$$\pi_{(3,\pm)}(\s) = \begin{pmatrix}
           0 & 1 & 0 \\ 0 & 0 & 1 \\ 1 & 0 & 0 
	 \end{pmatrix}, \quad
\pi_{(3,\pm)}(\t) = \pm \begin{pmatrix}
             1 & 0 & 0 \\ 0 & 0 & 1 \\ 0 & 1 & 0
	   \end{pmatrix}, $$
$$\pi_{(3,\pm)}((\s^i)^*)=0, \quad \pi_{(3,\pm)}((\s^i \t)^*)=I_3,\quad i =0,1,2.$$
\end{itemize}
\noindent Note that in each irrep, the non-zero dual elements correspond to precisely one conjugacy class of $D_3$.  Throughout this paper we denote the module associated with an irrep $\pi_\L$ by $V_\L$.

We utilise the $R$-matrices obtained in \cite{dil06}.  Explicitly we have the $R$-matrix $R_{(2,1)}(x) \in \End (V_{(2,1)} \otimes V_{(2,1)})$ given by 

\begin{equation*}
R_{(2,1)}(x) = \begin{pmatrix}
                  \o - x^2 & 0 & 0 & 0 \\
		  0 & -\o^{-1} (x^2-1)  &(\o-1)x  & 0 \\
		  0 & (\o - 1)x & -\o^{-1} (x^2-1) & 0 \\
		  0 & 0 & 0 & \o - x^2
		\end{pmatrix},
\end{equation*}

\noindent which is simply the six-vertex model at a cube root of unity.  Further, we have the $R$-matrix $R_{(3,+)} \in \End (V_{(3,+)} \otimes V_{(3,+)})$ given by 

\begin{equation*}
R_{(3,+)}(x)=
{\small \begin{pmatrix}
  x^2-x+1 & 0 & 0 & 0 & 0 & 0 & 0 & 0 & 0 \\
  0 & 0 & x(x-1) & x & 0 & 0 & 0 & 1-x & 0 \\
  0 & x(x-1) & 0 & 0 & 0 & 1-x & x & 0 & 0 \\
  0 & x & 0 & 0 & 0 & x(x-1) & 1-x & 0 & 0 \\
  0 & 0 & 0 & 0 & x^2-x+1 & 0 & 0 & 0 & 0 \\
  0 & 0 & 1-x & x(x-1) & 0 & 0 & 0 & x & 0 \\
  0 & 0 & x & 1-x & 0 & 0 & 0 & x(x-1) & 0 \\
  0 & 1-x & 0 & 0 & 0 & x & x(x-1) & 0 & 0 \\
  0 & 0 & 0 & 0 & 0 & 0 & 0 & 0  & x^2-x+1
      \end{pmatrix}.}
\end{equation*}

\section{Two universal Lax operators for $D(D_3)$}

\subsection{Universal Lax operator associated with a 2-dimensional representation}

A Lax operator $\mathfrak{L}(x) \in \text{End} (V_{(2,1)}) \otimes D(D_3)$ is given by 

\begin{equation*}
\mathfrak{L} (x)=\sum_{j \in \mathbb{Z}_3} 
	\begin{pmatrix}
		\o x \s^{-1} (\s^j \t)^* + (\o^j- \o x^2 c_1^2 \s^{-1})(\s^j)^* & 
			\o^j (\s^j \t)^* +  \frac13 (\o-1)c_1 x \o^j \s^j \t \s^* \\
		\o^j (\s^{-j} \t)^* + \frac13 (\o-1)c_1 x \o^j \s^{-j} \t (\s^{-1})^*  & 
			\o x\s (\s^j \t)^* + (\o^j -\o x^2 c_1^2 \s)(\s^{-j})^* 
	\end{pmatrix} \\
\end{equation*}

\noindent Here $c_1$ is the following Casimir element of $D(D_3)$:
$$c_1 = \frac 13 [(2e-\s-\s^{-1})(\s^* + (\s^{-1})^*].$$

\noindent  By direct calculation we have verified that $\mathfrak{L}(x)$ satisfies the algebraic relation \eqref{Lax} on $\End (V_{(2,1)} \otimes V_{(2,1)}) \otimes D(D_3)$ with the $R$-matrix $R_{(2,1)}(x)$, and is hence a univeral Lax operator.  In the limit $x\rightarrow 0$, this Lax operator reduces to $(\pi_{(2,1)} \otimes \text{id}) \mathcal{R}$ where $\mathcal{R}$ is the 
constant universal $R$-matrix given earlier.

Now defining $L_{(a,b)}(x)=(\text{id} \otimes \pi_{(a,b)}) \mathfrak{L} (x)$, we note that

\begin{align*}
L_{(2,e)}(x) &= I_2 \otimes I_2, \\ 
L_{(2,0)}(x) &= \text{diag}(\o, \o^{-1}, \o^{-1},\o), \\
L_{(2,1)}(x) &= R_{(2,1)}(x), \\
L_{(2,2)}(x) &= \text{diag}(\o-\o^{-1}x^2,\o^{-1}-x^2,\o^{-1}-x^2,\o-\o^{-1}x^2), \\ \\
L_{(3,\pm)}(x) &= 
	\begin{pmatrix}
	0 & 0 & \o x & 1 & 0 & 0 \\
	\o x & 0 & 0 & 0 & \o & 0 \\
	0 & \o x & 0 & 0 & 0 & \o^{-1} \\
	1 & 0 & 0 & 0 & \o x & 0 \\
	0 & \o^{-1} & 0 & 0 & 0 & \o x \\
	0 & 0 & \o & \o x & 0 & 0
	\end{pmatrix}
\end{align*}
It is straightforward to verify that these matrices satisfy the $RLL=LLR$ relation \eqref{RLL} 
on $V_{(2,1)} \otimes V_{(2,1)} \otimes V_\L$ for all 
irreducible modules $V_\L$, which again confirms that $\mathfrak{L}(x)$ is a universal Lax operator.

\subsection{Universal Lax operator associated with a 3-dimensional representation}

A Lax operator $\mathfrak{L}(x) \in \text{End} (V_{(3,+)}) \otimes D(D_3)$ is given by 

$$\mathfrak{L}(x) = \sum_{i,j = 1,2,3} E^i_j\otimes \left[(1-x)[((\s^{j-i})^* + (\s^{2-(i+j)}\t)^*] + x(x-1)c_1 \d^i_j \s^{i-1} \t 
+ x \s^{i-j} (\s^{i-1} \t)^* \right] .$$

\noindent In matrix form, this reads
{\scriptsize
\begin{multline*}
\mathfrak{L}(x) = \\
 \begin{pmatrix}
(1-x)[e^* + \t^*] + c_2 x(x-1)\t +x\t^* 
	& (1-x)[\s^* + (\s^{-1}\t)^*] +x\s^{-1}\t^* 
	& (1-x)[(\s^{-1})^*+(\s\t)^*]+x\s \t^* \\
(1-x)[(\s^{-1})^* + (\s^{-1}\t)^*]+x\s (\s\t)^* 
	& \hspace{-2mm}(1-x)[e^*+(\s\t)^*] + c_2x(x-1)\s\t +x(\s\t)^* 
	& (1-x) [\s^*+\t^*] + x \s^{-1}(\s\t)^* \\
(1-x)[\s^* + (\s\t)^*] +x \s^{-1}(\s^{-1}\t)^* 
	& (1-x) [(\s^{-1})^*+\t^*] + x\s(\s^{-1}\t)^*
	& \hspace{-2mm}(1-x)[e^* + (\s^{-1}\t)^*]+c_2 x(x-1)\s^{-1} \t +x(\s^{-1}\t)^*
\end{pmatrix}.
\end{multline*}
}

\noindent Here $c_2$ is the following Casimir element of $D(D_3)$:
%

$$c_2= \frac13 [2e-\s-\s^{-1}](\s^*+(\s^{-1})^*)+\sum_{k \in\mathbb{Z}_3}\s^k \t (\s^k\t)^*.$$
We have also verified that the above satisfies the defining relation \eqref{Lax} for a universal Lax operator.  Moreover, the expression again reduces to the constant solution of the Yang-Baxter equation in the limit $x \rightarrow 0.$  That is, $\lim_{x \rightarrow 0} \mathfrak{L}(x) = (\pi_{(3,+)} \otimes \text{id}) \mathcal{R}.$ 

 Setting $L_{(a,b)}(x) = (\text{id} \otimes \pi_{(a,b)}) \mathfrak{L}(x)$, we obtain

\begin{align*}
L_{(3, \pm)}(x) &= R_{(3,+)}(x) \\
L_{(2,e)}(x)&= (1-x)I_3 \otimes I_2, \\ 
L_{(2,j)}(x) & = (1-x) 
{\small \begin{pmatrix} 
0 & (1-\d^j_0) x &1 & 0 & 0 & 0\\
(1-\d^j_0) x & 0 & 0 & 0 & 0 & 1 \\
0 & 0 & 0 & (1-\d^j_0) x \o^{j} & 1 & 0 \\
0 & 1 & (1-\d^j_0) x \o^{-j} & 0 & 0 & 0 \\
1 & 0 & 0 & 0 & 0 & (1-\d^j_0) x \o^{-j}\\
0 & 0 & 0 & 1 & (1-\d^j_0) x \o^j & 0
\end{pmatrix}}, j \in \mathbb{Z}_3
\end{align*}
It can be verified that the 
above matrices satisfy the $RLL=LLR$ relation \eqref{RLL} on $V_{(3,+)} \otimes V_{(3,+)} \otimes V_\Lambda$ for all irreducible modules $\Lambda$.
~\\

\noindent
{\bf Acknowledgements} \\
We would like to thank Phil Isaac, Chris Campbell and Peter Finch for many helpful discussions.


\end{document}